\documentclass[intlimits,twoside,a4paper]{article}

\usepackage{amsmath,amssymb}
\usepackage{graphicx}

\usepackage[T2A]{fontenc}
\usepackage[cp1251]{inputenc}

\usepackage[eqsecnum]{cmpj2}

\usepackage{color}

\issue{2014}{17}{2}{23605}
\doinumber{10.5488/CMP.17.23605}

\title[Domain-walls in finite systems]%
{Domain-walls formation in binary nanoscopic finite systems}

\author[A. Patrykiejew, S. Soko{\l}owski]{A. Patrykiejew, S. Soko{\l}owski}
\address{Department for the Modelling of Physico-Chemical
Processes, Faculty of Chemistry, \\ Maria Curie-Sklodowska University, 20031 Lublin, Poland}

\authorcopyright{A. Patrykiejew, S. Soko{\l}owski, 2014}
\date{Received February 24, 2014, in final form May 6, 2014}

\begin{document}
\newcommand{\mtau}{\mbox{\boldmath$\tau$}}
\newcommand{\mqu}{\mbox{\bf q}}
\newcommand{\mbr}{\mbox{\bf r}}
\newcommand{\mbR}{\mbox{\boldmath$R$}}

\maketitle

\begin{abstract}
Using a simple one-dimensional Frenkel-Kontorowa type model, we have demonstrated that
finite commensurate
chains may undergo the commensurate-incommensurate
(C-IC)
transition when the
chain is contaminated by isolated impurities attached to the chain ends.  Monte Carlo
(MC)
simulation has
shown that the same phenomenon appears in two-dimensional systems with impurities located at the
peripheries of finite commensurate clusters.

\keywords{binary mixture, commensurate-incommensurate transitions, Monte Carlo simulation, finite systems, Frenkel-Kontorova model}
\pacs{64.70.Rh, 68.55.Ln, 64.60.an }

\end{abstract}

\section{Introduction}
In modern nanotechnologies one often deals with very small
systems of countable numbers of atoms
or molecules. In such cases,
the  finite size and boundary effects are  large and bound to significantly affect the properties of
the system with respect to its bulk counterpart \cite{nano1}. Another
important problem is the purity of
small systems. While
tiny amounts of impurities may be unimportant in macro-scale, the behavior of nanoscopic
systems is considerably influenced even by a small number of impurity atoms \cite{imp1}.
Among the systems in which the presence of impurities may be of importance are those exhibiting the
C-IC transition.

The C-IC transitions have been experimentally observed
in a variety of systems including adsorbed
films, \cite{cicex1,cicex2,cicex4,cicex8}
intercalated compounds \cite{cicex9,cicex11}
composite crystals \cite{cicex13}
and
magnetically ordered structures of rare-earth compounds \cite{cicex15}.
Theoretical studies of C-IC transitions have focused on
the domain wall description of incommensurate
phases \cite{cic2,cic4,cic6,cic7}.
According to the
domain wall formalism, the IC phase is a collection of
C domains separated by domain walls.
The density within the domain walls may be
lower or higher than the density of commensurate domains. In the former
case, the walls are light and superlight, while in the latter the walls are
heavy and superheavy \cite{cic6}.

The simplest theoretical approach which predicts the formation of domain walls is
the one-dimen\-sional Frenkel-Kontorova model \cite{FK1}.
The original FK model assumes that an infinite chain
of atoms interacting via harmonic potential at zero temperature is subjected
to a periodic (sinusoidal) external field.
Depending on the misfit between the equilibrium distance of the harmonic potential,
the period and amplitude
 of the external field, the FK model is capable of
describing the C-IC transition. The FK model has been extended
to two-dimensional systems \cite{FK2D-1,FK2D-2} to
mixtures \cite{FK-mix1}, systems with disorder \cite{FK-dis2}
and has also been used to study finite chains \cite{FK-fin2,FK-fin3,FK-fin4}.

The NPT Monte Carlo
simulation has demonstrated \cite{phillips2005,phillips2008} that
finite one-dimensional chains,
either uniform or subjected to periodic
field,
exhibit structures that cannot
appear in
infinite chains.  In particular, it has been shown that the chain
experiences very large density fluctuations.
In the case of chains on a periodic substrate, a number of
different structures (registered, free floating, domain-wall
incommensurate and resulting from the chain fragmentation)
have been found to appear during a single run.

In one of our recent papers \cite{myArKr}, we have shown that
finite two-dimensional clusters of Kr adsorbed
on graphite undergo the
C-IC phase transition
when contaminated by small amounts of Ar atoms.
Computer simulation has demonstrated that the transition
occurs already when the boundaries of a finite krypton island
are covered with a single layer of argon atoms.

In this paper we address the issue of the influence of  impurities on the behavior of
finite one- and two-dimensional (1D and 2D) systems.
We are interested in the effects of impurities located at the
peripheries of finite 1D
chains and
2D clusters of atoms subjected to the periodic
external field. The field is assumed to be strong enough to enforce
the formation of commensurate structures
in pure systems and we consider
the possibility of the
C-IC transition driven
by the presence of impurities.  The paper is organized as follows.
In the next section we discuss the
behavior of one-dimensional systems in the framework of a modified
Frenkel-Kontorova model. Then, in the third section
we consider two-dimensional finite systems studied by Monte Carlo simulation.
In the final section we summarize our findings.

\section{One-dimensional Frenkel-Kontorova model}

 At first, we have considered a simple
 1D finite chain of atoms at zero temperature and
used the
Frenkel-Kontorova (FK) model \cite{FK1}. The energy of a finite chain
consisting of $N$ atoms subjected
to a periodic external potential and containing impurities can be written as follows:
\begin{equation}
E = \frac{1}{2}\left\{\sum_{i=1}^{N-1}K_{i}[x_{i+1}-x_{i}-b_{i}]^2 + \sum_{i=1}^Nv_{i}[1-\cos(2\pi x_i/a)]\right\},
\label{eq:Kon1}
\end{equation}
where $K_{i}$ and $b_{i}$ are the elastic constant and the equilibrium distance for the pair
$(i,i+1)$ and $v_i$ is the amplitude of the external field for the $i$-th particle.
Having introduced the displacements
$u_i = x_i/a-pi$ ($i=1,2,\ldots N$) with $p$ being a positive integer
(in this work we set
$p=2$),
the energy given by eqn.(\ref{eq:Kon1})
can be rewritten in units of  $K_{0}a^2/2$ ($K_{0}$ being the elastic constant for a pure chain) as follows:
\begin{equation}
E=\sum_{i=1}^{N-1}\hat{K}_{i}[u_{i+1}-u_{i}-m_{i}]^2 + \sum_{i=1}^N\hat{v}_{i}[1-\cos(2\pi u_i)].
\label{eq:Kon2}
\end{equation}

In general, the elastic constants $\hat{K}_{i}$ (the misfits $m_{i}$) can assume one
of three possible values $k_{0,0}\equiv 1$,  $k_{0,\textrm{im}}$
or $k_{\textrm{im},\textrm{im}}$ ($m_{0,0}$, $m_{0,\textrm{im}}$ or $m_{\textrm{im},\textrm{im}}$)
depending on the composition of the pair $(i,i+1)$, and
the amplitude $\hat{v}_i = v_i/a^2K_{0}$ is equal either to $v_0$ or to $v_\textrm{im}$.

In order to find the equilibrium configuration of the chain, the energy
should attain
its
minimum value, specified by the condition stating that the forces
$f_i=-\partial E/\partial u_i = 0$ for all $i$.
We consider the systems
in which a single impurity atom is located
at one end of the chain (class I) and the systems with two impurity atoms
located at both ends of the chain (class II) and put $k_{0,\textrm{im}}=k_\textrm{im}$ and $m_{0,\textrm{im}}=m_\textrm{im}$.
The behavior of pure chains depends on $m_{0}$ and $v_0$. For the assumed
value of $m_{0}=-0.1$, pure chains are commensurate when $v_0$ exceeds the critical value
$v_{0,\textrm{c}}\approx 0.004950$. The calculations
have been done for  $N$ between 21 and 401 and $v_0=0.006$.

\begin{figure}[!t]
\centerline{
\includegraphics[width=0.6\textwidth]{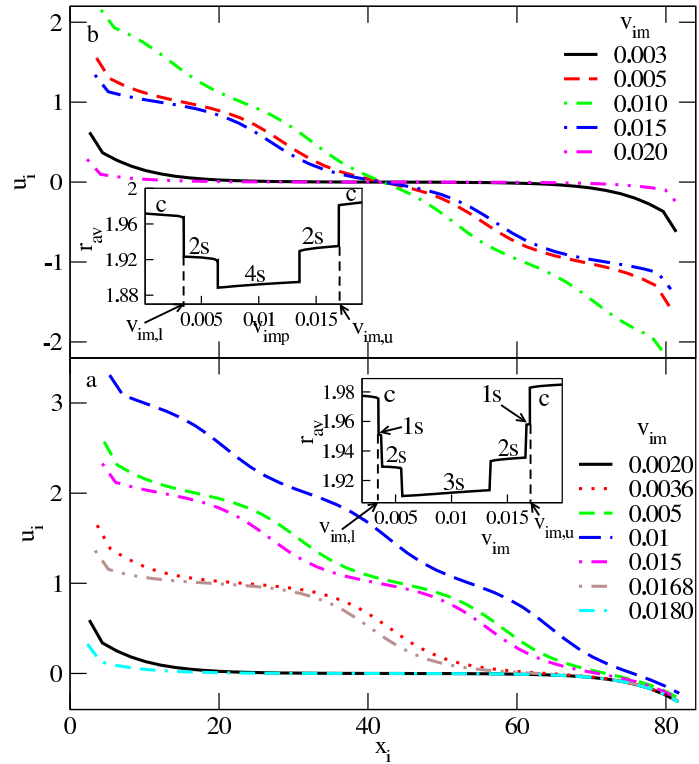}          
}
\caption{(Color online) Atomic displacements vs.
atomic positions for the systems
of class I (part a) and II (part b) and different values of $v_\textrm{im}$ (given in the figure).
The calculations have been done
$N=41$, $m_\textrm{im}=-0.25$ and $k_\textrm{im}=1.0$. The insets show the changes of the average
nearest-neighbor distance
vs. $v_\textrm{im}$ The regions marked by $C$ and $ks$ ($k=1,2,3,4$) correspond to the
commensurate structure to the incommensurate structures with $k$ domain walls, respectively.}
\label{fig1}
\end{figure}

The first series of calculations have
been carried out for $k_\textrm{im}=1.0$,
while  $v_\textrm{im}$ and $m_\textrm{im}$  have been varied.
Figure~\ref{fig1} shows the example of the results
obtained for the chains with $N=41$, $m_\textrm{im}=-0.25$ and different
values of $v_\textrm{im}$. The lower and upper panels of figure~\ref{fig1} show the results for the
systems of class I and II,
respectively.  It is
evident that the systems belonging to both
classes exhibit a qualitatively similar
behavior. Of course, the systems of the class I
lack
the symmetry of atomic displacements with respect to the central atom.
For $v_\textrm{im}$ lower (higher)
than $v_{\textrm{im},\textrm{l}}$ ($v_{\textrm{im},\textrm{u}}$), the chain remains commensurate,
while for intermediate values of $v_\textrm{im}$, there appear incommensurate structures with domain walls.
When $v_\textrm{im}$ is lower than
$v_{\textrm{im},\textrm{l}}$, the energy cost to put the impurity out of registry position is low and  the impurity
can exhibit large displacements from the commensurate position, while the rest of the
chain assumes a commensurate structure due to the domination of the surface over the
elastic interaction. On the other hand, when $v_\textrm{im}>v_{\textrm{im},\textrm{u}}$, the impurity is
strongly pinned by the surface
potential and the chain retains the commensurate structure.
Figure~\ref{fig1} shows that when $v_\textrm{im}>v_{\textrm{im},\textrm{u}}$,
the displacements of impurities are considerably lower than
when $v_\textrm{im}<v_{\textrm{im},\textrm{l}}$.

For intermediate values of $v_\textrm{im}$, the gain in elastic
energy due to transition into the incommensurate structure
is larger than the loss of surface energy. The insets to figure~\ref{fig1}
show the changes of the average nearest-neighbor
distance with $v_\textrm{im}$. In both classes of systems, we have
found a series of transitions characterized by a different
number of domain walls. The calculations for chains with $N$
up to 401 have shown that in longer chains a larger number of
structures appear, and it seems that in the limit of $N\rightarrow\infty$ the transitions
form a harmless staircase  \cite{cic2}.

\begin{figure}[!t]
\centerline{
\includegraphics[width=0.6\textwidth]{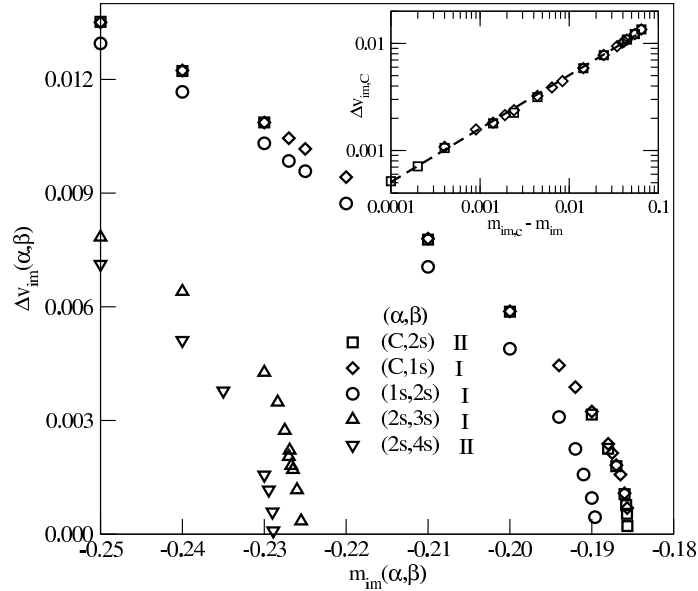}
}
\caption{The phase diagrams showing the dependence of the difference
between the upper and lower values of $v_\textrm{im}$
at the transition points ($\Delta v_\textrm{im}(\alpha,\beta)$) between different structures ($\alpha,\beta$)
and the impurity misfit  $m_\textrm{im}(\alpha,\beta)$.  The calculations have been done
for  $N=41$ and $k_\textrm{im}=1.0$ and for the classes I and II.
The inset shows the scaling plot for the C-IC transition.}
\label{fig2}
\end{figure}

The values of $v_{\textrm{im},\textrm{l}}$ and $v_{\textrm{im},\textrm{u}}$ change with $m_\textrm{im}$ and there is a critical value
of $m_{\textrm{im},\textrm{c}}$ for which
the difference $\Delta v_\textrm{im} = v_{\textrm{im},\textrm{u}}-v_{\textrm{im},\textrm{l}}$ goes to zero (figure~\ref{fig2}).
In the particular examples
considered here ($N=41$ and $k_\textrm{im}=1.0$),  $m_{\textrm{im},\textrm{c}}\approx -0.1856$
for the systems of class I and II.
The inset to figure~\ref{fig2} shows that $\Delta v_\textrm{im}$ scales with
$m_{\textrm{im},\textrm{c}} - m_\textrm{im}$ as follows:
\begin{equation}
\Delta v_\textrm{im}\propto (m_{\textrm{im},\textrm{c}} - m_\textrm{im})^{1/2}, \qquad k=1,2.
\label{eq:Kon10}
\end{equation}
The same scaling
appears for the transitions between different incommensurate structures, but
the value of $m_{\textrm{im},\textrm{c}}$ for each transition is different (figure~\ref{fig2}).

\begin{figure}[!b]
\centerline{
\includegraphics[width=0.6\textwidth]{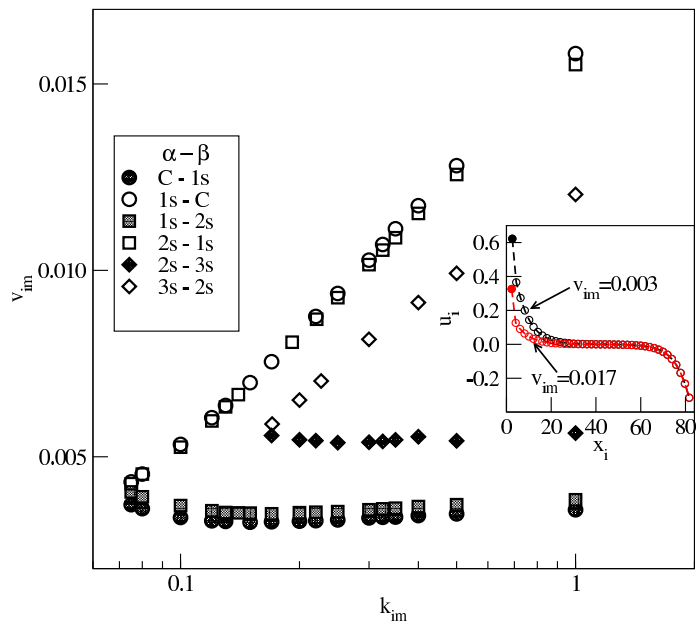}
}
\caption{(Color online) The phase diagrams showing the dependence of  $v_\textrm{im}$ on $k_\textrm{im}$
at the transition points between different structures.  The calculations
have been done for  $N=41$ and $m_\textrm{im}=-0.24$.
The inset shows the atomic displacements versus actual atomic positions in the commensurate phase
for  $v_\textrm{im}=0.003$ and $v_\textrm{im}=0.017$  when $k_\textrm{im}=1.0$. Filled symbols
mark the impurity atom.}
\label{fig3}
\end{figure}

We have then investigated the effects due to changes in the magnitude of $k_\textrm{im}$.  Figure~\ref{fig3} gives
the example of results
for the systems
of class I.
We see that the values of $v_\textrm{im}$ at the transitions $C-1s$, $1s-2s$ and $2s-3s$
are nearly independent of $k_\textrm{im}$, while those corresponding to
transitions $3s-2s$, $2s-1s$ and  $1s-C$ (leading to the recovery
of commensurability), exhibit a logarithmic dependence on $k_\textrm{im}$.

The logarithmic dependence of $v_\textrm{im}$ on $k_\textrm{im}$ at transitions $3s-2s$, $2s-1s$ and  $1s-C$
has the same origin as the
C-IC transition in a pure FK model \cite{cic2}.
On the other hand, the mechanism of transitions leading
to commensurability when $v_\textrm{im}$ becomes very low is different. For sufficiently
low $v_\textrm{im}$, the energy cost to put the impurity out of registry position and
to restore the
commensurate positions in the rest of the chain is low. Consequently, the impurity
can exhibit a large displacement from the commensurate position, while the rest of the
chain assumes a commensurate structure due to the domination of surface energy over the
elastic energy. This is illustrated in the inset to figure~\ref{fig3}, which shows atomic displacements for the
systems with $k_\textrm{im}=1.0$ and $v_\textrm{im}= 0.003$ and $v_\textrm{im}=0.017$, while the rest of the
parameters have been kept
the same as in the main figure. Of course, there is an asymmetry of displacements in
the chain but it is very small. In the commensurate phase at
high values of $v_\textrm{im}=0.017$, the displacements
at both ends are nearly the same.

Another question is whether a single impurity can drive the incommensurate
system into commensurability?
The answer is no. On the other hand, two impurities located at both ends of
the chains do lead to the recovery of commensurability when $v_\textrm{im}$ is sufficiently high.
This is just the same as with the rope pinned either to one or two walls. In the first
case, the rope hangs down freely. In the latter, the rope pinned to the opposite walls
can be expanded to some extent. In the harmonic approximation, the chain always
retains integrity, although when the interaction potential allows for dissociation,
the chain may rupture \cite{markov,phillips2005,phillips2008} rather than restore the C structure.

\section{Two-dimensional finite clusters}

The phenomenon of reentrant commensurability is not restricted
to the above discussed simple 1D model.
It also appears in more realistic 2D systems in the ground state and at finite temperatures.
To demonstrate
this, we have performed
MC simulation in the canonical ensemble for finite clusters decorated with
impurities at the boundaries. The particles of $A$ and $B$ (impurity) have been placed
in the substrate field
\begin{equation}
v_i(x,y)=-V_{i}\{\cos({\mqu}_1{\mbr})+\cos({\mqu}_2{\mbr})+\cos[({\mqu}_1-{\mqu}_2){\mbr}]\}
\label{eq:graphite}
\end{equation}
corresponding to the graphite basal plane,
where $V_i$ is the amplitude of the potential for the $i$th
 component and ${\mqu}_1$ and ${\mqu}_2$
are the reciprocal lattice vectors of the graphite basal plane. The initial configurations
have been
a hexagon of a perfectly ordered $(\sqrt{3}\times\!\!\sqrt{3})R30^\circ$ structure consisting
of $N_Az$ atoms of $A$ and $N_B$ atoms of $B$ placed along the one edge (case I), the adjacent
three edges (case II) or along
all six edges (case III) of the cluster.
The interaction
between the particles has been
represented by the LJ(12,6)
potential with the fixed $\sigma_{AA}=1.46a$ ($a=2.46$~{\AA}  is the graphite lattice constant
taken as the unit of length) and
$\varepsilon_{AA}/k_\textrm{B}=170$~K (assumed to be a unit of energy) and
$V^{\ast}_A=V_A/\varepsilon_{AA}= 0.12$.
In order to reduce the tendency of $B$ atoms towards aggregation, we have put
$\varepsilon_{BB}=0.5\varepsilon_{AA}$, while  $\varepsilon_{AB}$ has been assumed to be equal
to $\varepsilon_{AA}$. The parameters  $\sigma_{BB}=\sigma_{AB}$ and $V_B$ have been varied.

The simulations have been carried out for the systems with $N_A= 271$ and $N_B=11$
(case I), $N_B=33$ (case II) and
$N_B=66$ (case III). Besides, we have also performed some runs for a larger system
corresponding to case III, with
$N_A=811$ and $N_B=114$ as well as with $N_A=1189$ and $N_B=138$. The simulation has
been performed at reduced temperatures $T^{\ast}=k_\textrm{B}T/\varepsilon_{AA}$ between 0.005 and 0.3.
The formation of domain walls in the system has been monitored using the
order parameter \cite{myArKr,HoulLan}
\begin{equation}
\phi({\mbr})=\cos({\mqu}_1{\mbr})+\cos({\mqu}_2{\mbr})+\cos[({\mqu}_1-{\mqu}_2){\mbr}],
\label{eq:local}
\end{equation}
and assuming that the atom is commensurate (incommensurate) when $\phi>0$ ($\phi\leqslant 0$).
We have calculated
the average numbers of incommensurate atoms $A$ and $B$ neglecting the atoms located
at the patch boundaries, i.e., those with less than 5 nearest neighbors.

It has been found
that for any
value of $\sigma^{\ast}_{BB}=\sigma_{BB}/a$ between 1.34 and 1.37 and for
sufficiently low and sufficiently high values of $V^{\ast}_B$, the systems
remain commensurate,
while for the intermediate values of $V^{\ast}_B$, there appear networks of heavy walls
separating commensurate domains.

\begin{figure}[!t]
\centerline{
\includegraphics[width=0.7\textwidth]{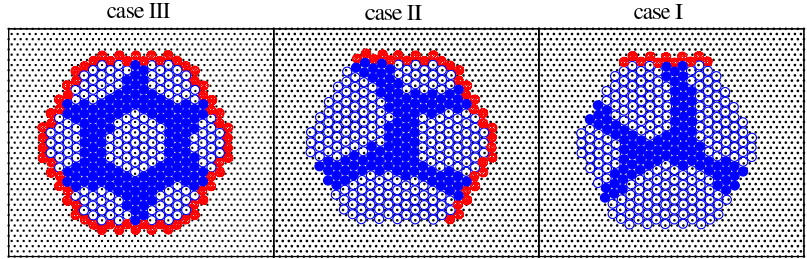}
}
\caption{(Color online) The examples of snapshots  for the system with  $\sigma^{\ast}_{BB} =1.36$
and $V^{\ast}_B =0.12$ at $T^{\ast}= 0.02$ corresponding to the three cases considered.
Shaded circles represent $B$ atoms. Open (filled) circles correspond
to the commensurate (incommensurate) $A$ atoms.}
\label{fig4}
\end{figure}

In general, the results at low temperatures are qualitatively very similar to
those obtained for 1D FK model.
In particular, we have observed the formation of IC structures
in the cases I, II and III, though the structure of the domain-wall networks is different in
all cases. In general, the walls tend to assume the orientations perpendicular to the edges
decorated with the impurity atoms. Therefore, in case III, we have found
regular networks like that given in
the leftmost part of figure~\ref{fig4}. On the other hand, the systems with only one
or three adjacent edges decorated by impurities form irregular networks of
domain walls (see the middle and rightmost panels to figure~\ref{fig4}).

\begin{figure}[!b]
\centerline{
\includegraphics[width=0.48\textwidth]{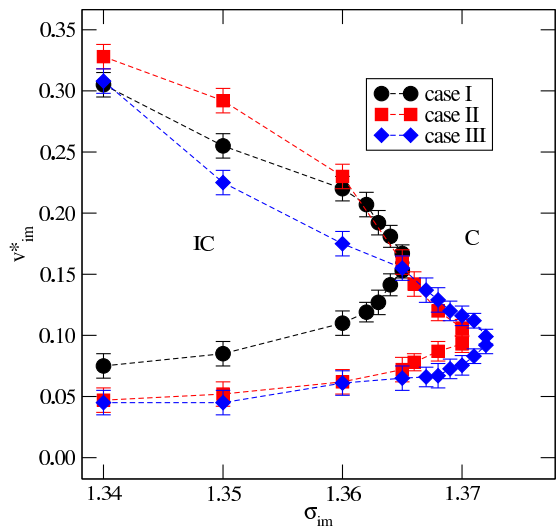}
}
\caption{(Color online) The phase diagram showing the locations ($V^{\ast}_B$ versus $\sigma^{\ast}_{AB}$)
of C-IC transition in the limit of $T^{\ast}\rightarrow 0$
in the systems with $N_A=271$ and different concentration of impurity atoms.}
\label{fig5}
\end{figure}

\begin{figure}[!t]
\vspace{4mm}
\centerline{
\includegraphics[width=0.5\textwidth]{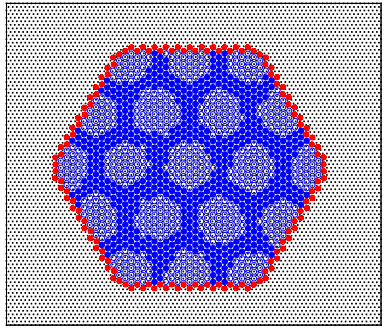}
}
\caption{(Color online) The example of snapshot  for the system
with $N_A=1189$ and $N_B=138$, $\sigma_{AB} =1.365$
and $V^{\ast}_B=0.08$ recorded at $T^{\ast}=0.01$~K. Shaded circles represent $B$ atoms.
Open and filled circles
correspond to the commensurate and incommensurate $A$ atoms.}
\label{fig6}
\end{figure}

The region of $V^{\ast}_B$ over
which the IC phase is stable at low temperatures gradually decreases when the size
of impurity atoms increases, so that for sufficiently large impurity atoms, only the
C phase occurs
independently of the magnitude of $V^{\ast}_B$ (see figure~\ref{fig5}). The calculations
for the larger system
with $N_A = 811$ and $N_B=114$ (case III)  have shown that  maximum value of
$\sigma^{\ast}_{AB}$ above which the C-IC transition
does not appear slightly increases with the system size. Moreover,
the simulations performed for larger systems  have also demonstrated that at
low temperatures the size of C domains does not change and only their number increases. This
result is 
quite similar to that obtained for 1D FK model for the chains of different length.
This is demonstrated by the snapshots given in the leftmost part of figure~\ref{fig4} and in figure~\ref{fig6}.

The impurity induced changes of the surface stress are responsible for the C-IC transition,
which leads to the compression of the finite island. When the amplitude of the
impurity surface potential, $V^{\ast}_B$, is very low, the
presence of impurities at the cluster boundaries does not
appreciably affect the strain of the atoms in the cluster. Therefore, the surface
stress of the impurity decorated
island does not match differently
from that of the pure system. For
 sufficiently large values of $V^{\ast}_B$, the impurities are strongly pinned over the surface
potential minima as well as do not enlarge the strain in the island. Therefore,
for sufficiently low and
high $V^{\ast}_B$, the system retains the C structure. For intermediate values of $V^{\ast}_B$,
the competing atom-atom and atom-external field interactions lead to a sufficiently
large increase of a surface stress in order to
trigger the C-IC transition.
One should note a similarity of the herein reported impurity driven C-IC transition
to the behavior of metal nanowires
 \cite{nwir1} and to the formation of IC phases in the subjected to uniaxial compression
self-assembled monolayer gold nanoparticles
supported on a fluid  \cite{gold}.

The results presented here describe only the low temperature behavior of the systems.
The
simulation has demonstrated that upon an increase of temperature,
the IC structure gradually changes and
the C structure is restored at temperatures still below the melting point
of the cluster. The
effects of thermal excitations on the domain-wall networks will be discussed elsewhere.

 \section{Summary}

In this work we have studied the impurity driven commensurate-incommensurate
transitions in one-  and two-dimensional
finite systems at zero temperature. In the case of one-dimensional finite chains,
we have considered the situations in which the impurity is located at one and two ends of the chain.
It has been shown that in both situations, the C-IC transition occurs when the
amplitude of the external field experienced by the impurity atoms falls into the region between the
lower and upper threshold values. These limiting values of $v_\textrm{im}$ depend upon
the parameters characterizing the interaction between the atoms in the main chain,
the amplitude of external field acting on the main chain atoms
and the interaction between the main chain and the impurities. The number of solitons
(domain walls) in the IC structure is different for the chains with one and two
ends decorated with the impurity atoms and also depends upon the $v_\textrm{im}$.
This behavior is a consequence of different symmetry properties of the system.

In the case of two-dimensional finite clusters, a very similar C-IC transition
has been found. In particular, we have observed that the transition occurs
when only a part of the cluster boundary is covered with a single row of impurity atoms.
Also, by analogy to the results of one-dimensional FK model, the transition
occurs only between the lower and upper threshold values of $v_\textrm{im}$.

It should be noted that the C-IC transition observed in two-dimensional systems  occurs only
when the parameters entering the potential describing the interactions between  $AA$, $AB$,
and $BB$ atoms are suitably chosen. First of all, the components $A$ and $B$ should not exhibit the
tendency towards mixing. Otherwise, the impurity atoms $B$ are likely to penetrate the patch
and different scenarios are possible. One, is the formation of a mixed commensurate phase. This
has been found in the case of Ar-Kr finite patches at sufficiently high temperatures \cite{myArKr}, when
the $AB$ interaction potential parameters were obtained using the standard Lorentz-Berthelot
mixing rules. The same mixture exhibits a different behavior at low temperatures, and does undergo the
C-IC transition, although with the domain walls preferentially made of Ar atoms.
Another requirement is that atoms $A$ should tend to order into the C phase, while atoms $B$ should order into the IC
phase. This imposes certain restrictions on the choice of $\sigma_{AA}^{\ast}$, $\sigma_{BB}$ and the
values of $V_A^{\ast}$ and $V_{B}^{\ast}$. In particular, $\sigma_{AA}^{\ast}$ and $V_A^{\ast}$ should
assume the values ensuring that a pure $A$ patch is commensurate, but are likely to undergo the C-IC transition
when the density exceeds the monolayer capacity or when subjected to an external force due to the
presence of impurity atoms along the patch boundaries. Thus, $V_{A}^{\ast}$ cannot be too low or
to high and the misfit between the sizes of surface lattice and adsorbate atoms is rather small. A good
example of such a system is krypton adsorbed on graphite \cite{cicex4}. In fact, the parameters  $\sigma_{AA}^{\ast}$ and $V_A^{\ast}$ used here are rather close to those describing the krypton
adsorbed on the graphite basal plane \cite{myArKr}. On the other hand, the values of $\sigma_{BB}$
and $V_{B}^{\ast}$ should favor the formation of IC phase by pure component $B$. Thus, the misfit
between the size of B atoms and the surface lattice should be sufficiently large (it can be positive as well as
negative) and $V_{B}^{\ast}$ should be sufficiently small. However, there is still another requirement for the
appearance of C-IC transition in finite patches. The $AB$ interaction should be sufficiently strong so that
the layer of $B$ atoms along the patch boundary is stable. Otherwise, the atoms $B$ would  prefer
to form clusters rather than stay at the patch boundary. Besides, this condition is also necessary to
exert a sufficiently large force upon the atoms $A$ inside the finite patch to trigger the C-IC transition.
Usually, the C-IC transition occurs only when the film density exceeds the density of a fully filled
commensurate phase. For the C-IC transition to occur in finite patches, the stability of the C phase should
be low enough so that a rather small force exerted by a thin layer of impurity atoms along the patch
boundary is able to drive the C-IC transition. From our earlier studies of adsorption of Ar-Kr,
Ar-Xe and Kr-Xe mixtures \cite{myArKr,myArXe,myKrXe,mymix} on graphite it follows that only in the case of
Ar-Kr mixture the C-IC transition occurs at submonolayer densities in finite patches of adsorbed phase.

One expects that at higher temperatures, thermal excitations may considerably influence the effects observed.
This problem is currently under study and the results will be published elsewhere.

\noindent
\section*{Acknowledgements}

\noindent
This work
 was  supported by ERA under the Grant
 PIRSES 268498.

\ukrainianpart

\title{Формування домен-стінки в бінарних наноскопічних скінченних системах}

\author{А. Патрикєєв,
        С. Соколовскі}
\address{Відділ моделювання фізико-хімічних процесів, Університет Марії Кюрі-Склодовської, Люблін, Польща}

\makeukrtitle

\begin{abstract}
\tolerance=3000%
Використовуючи просту одновимірну типову модель Френкеля-Конторової, ми показали, що скінченні співвимірні ланцюжки
можуть зазнавати співвимірний-неспіввимірний (C-IC) перехід, якщо ланцюжок забруднений ізольованими
домішками, дочепленими до кінців ланцюжка.  За допомогою методу Монте Карло (MC) показано,
що таке ж явище виникає в двовомірних системах з домішками, розміщеними на периферії скінчених співвимірних кластерів.

\keywords{бінарна суміш, співвимірні-неспіввимірні переходи, симуляції Монте Карло, скінченні системи, модель Френкеля-Конторової
}
\end{abstract}


\begin{thebibliography}{99}
\bibitem{nano1}
{Esfarjani~K., Mansoori~G.A.,  In:
{Handbook of Theoretical and Computational Nanotechnology},
Vol.~10, Rieth M., Schommers W. (Eds.), American Scientific Publishers, Los Angeles, 2005, 1--45.}

\bibitem{imp1} Hwang~I.-S.,  Fang~C.-K., Chang~S.-H., {Phys. Rev. B},  2011,
 \textbf{83}, 134119.\; \bibdoi{10.1103/PhysRevB.83.134119}.



\bibitem{cicex1} Jaubert~M., Glachant~A., Bienfait~M., Boato~G.,
{Phys. Rev. Lett.}, 1981, \textbf{46}, 1679; \bibdoi{10.1103/PhysRevLett.46.1679}.

\bibitem{cicex2} Krim~J., Suzanne~J., Shechter~H., Wang~R.,
Taub~H., {Surface Sci.}, 1985, \textbf{162}, 446; \doi{10.1016/0039-6028(85)90933-1}.



\bibitem{cicex4}  {Stephens~P.W.,  Heiney~P.A., Birgeneau~R.J.,
Horn~P.M.,  Moncton~D.E.,  Brown~G.S., {Phys. Rev. B},
1984, \textbf{29}, 3512; \\ \bibdoi{10.1103/PhysRevB.29.3512}.}


\bibitem{cicex8} Usachov~D.,  Dobrotvorskii~A.M., Varykhalov~A.,
 Rader~O., {Phys. Rev. B}, 2008, \textbf{78}, 085403; \\ \bibdoi{10.1103/PhysRevB.78.085403}.

\bibitem{cicex9} {Clarke~R., In: {Ordering in Two Dimensions}, Sinha S.K. (Ed.),
 North-Holland, Amsterdam, 1980, p. 53--58.}


\bibitem{cicex11} Li~L.J., Lu~W.J.,  Zhu~X.D.,  Ling~L.S.,
Qu~Z.,  Sun~Y.P.,  {Europhys. Lett.}, 2012, \textbf{97}, 67005; \\ \doi{10.1209/0295-5075/97/67005}.


\bibitem{cicex13} Nuss~J., Pfeiffer~S., van Smallen~S.,
Jansen~M.,  {Acta Crystalogr. B}, 2010, \textbf{66}, 27; \doi{10.1107/S0108768109053312}.


\bibitem{cicex15} Vokhmyanin~A.P., Lee~S.,
Jang~K.-H.,  Podlesnyak~A.A.,  Keller~L., Proke\v{s} K.,
Sikolenko~V.V., Park~J.-G., Skryabin~Yu.N.,
Pirogov~A.N., {J. Magn. Magn. Mater.}, 2006, \textbf{300}, e411; \bibdoi{10.1016/j.jmmm.2005.10.179}.


\bibitem{cic2} Bak~P., {Rep. Prog. Phys.}, 1982, \textbf{45}, 587; \doi{10.1088/0034-4885/45/6/001}.


\bibitem{cic4} Huse~D.A.,  Fisher~M.,  {Phys. Rev. B},
1984, \textbf{29}, 239; \bibdoi{10.1103/PhysRevB.29.239}.


\bibitem{cic6} {Den Nijs~M., In: {Phase Transitions and Critical Phenomena}, Vol.~12,
Domb~C., Lebowitz J.L. (Eds.), Acedemic Press, London, 1988, p. 219--333.}

\bibitem{cic7} Patrykiejew~A., Soko{\l}owski~S.,
Binder~K., {Surface Sci.}, 2002, \textbf{512}, 1; \bibdoi{10.1016/S0039-6028(02)01702-8}.


\bibitem{FK1} Frenkel~Y.I.,  Kontorova~T.,
{Zh. Eksp. Theor. Fiz.}, 1938, \textbf{8}, 1340 (in Russian).

\bibitem{FK2D-1}  Lomdahl~P.S.,  Srolovitz~D.J.,
{Phys. Rev. Lett.}, 1986, \textbf{57}, 2702; \bibdoi{10.1103/PhysRevLett.57.2702}.

\bibitem{FK2D-2} Hamilton~J.C.,  {Phys. Rev. Lett.}, 2002, \textbf{88}, 126101; \bibdoi{10.1103/PhysRevLett.88.126101}.

\bibitem{FK-mix1} Daruka~I., Hamilton~J.C.,
{J. Phys.: Condens. Matter}, 2003, \textbf{15}, 1827; \bibdoi{10.1088/0953-8984/15/12/302}.

\bibitem{FK-dis2}  Braun~O.M.,  Kivshar~J.S., {Phys. Rep.},
1998, \textbf{306}, 1; \bibdoi{10.1016/S0370-1573(98)00029-5}.


\bibitem{FK-fin2}  Braiman~Y., Baumgarten~J., Jortner~J., Klafter~J.,
{Phys. Rev. Lett.},  1990, \textbf{65}, 2398; \bibdoi{10.1103/PhysRevLett.65.2398}.

\bibitem{FK-fin3}  Braiman~Y., Baumgarten~J.
Klafter~J., {Phys. Rev. B}, 1993, \textbf{47}, 11159; \bibdoi{10.1103/PhysRevB.47.11159}.

\bibitem{FK-fin4} Vanossi~A., Franchini~A.,
Bortolani~V.,  {Surface Sci.}, 2002, \textbf{502-503}, 437; \bibdoi{10.1016/S0039-6028(01)01990-2}.


\bibitem{phillips2005}  Phillips~J.M., Dash~J.G.,
{J. Stat. Phys.}, 2005, \textbf{120},  721; \doi{10.1007/s10955-005-5252-x}.


\bibitem{phillips2008} Hartnett~A.S.,  Phillips~J.M.,
{Phys. Rev. B},  2008, \textbf{77}, 035408; \bibdoi{10.1103/PhysRevB.77.035408}.


\bibitem{myArKr} Patrykiejew~A., R\.zysko~W., Soko{\l}owski~S.,
{ J. Phys. Chem. C}, 2011, \textbf{116}, 753; \bibdoi{10.1021/jp208323b}.


\bibitem{markov} Markov~I., Trayanov~A.,
{ J. Phys.: Condens. Matter}, 1990, \textbf{2}, 6965; \bibdoi{10.1088/0953-8984/2/33/009}.

\bibitem{HoulLan} Houlrik~J. M., Landau~D.P.,
{ Phys. Rev. B},  1991, \textbf{44}, 8962; \bibdoi{10.1103/PhysRevB.44.8962}.

\bibitem{nwir1} Diao~J., Gall~K., Dunn~M.L.,
{Nat. Mater.}, 2003, \textbf{2}, 656; \bibdoi{10.1038/nmat977}.

\bibitem{gold} {Chua~Y., Leahy~B., Zhang~M.,
You~S.,  Lee~K.Y.C.,   Coppersmith~S.N.,  Lin~B.,
{PNAS},  2013, \textbf{110}, 824;\\
\bibdoi{10.1073/pnas.1101630108}.}

\bibitem{myArXe} Patrykiejew~A., {Condens. Matter Phys.}, 2012, \textbf{15}, 23601. \bibdoi{10.5488/CMP.15.23601}.

\bibitem{myKrXe}Patrykiejew~A., Soko{\l}owski~S.,
{ J. Chem. Phys.}, 2012, \textbf{136}, 144702; \bibdoi{10.1063/1.3699330}.

\bibitem{mymix} Patrykiejew~A., {J. Phys.: Condens. Matter}, 2013, \textbf{25}, 015001; \bibdoi{10.1088/0953-8984/25/1/015001}.

\end{thebibliography}
\end{document}